# Determining Meteorological Tidal Transport through a Channel on the Coast

Chunyan Li[a,b]

[a] *Department of Oceanography and Coastal Sciences, Louisiana State University, Baton Rouge, LA, USA*
[b] *Coastal Studies Institute, Louisiana State University, Baton Rouge, LA, USA*

Abstract.

Transient weather systems are often associated with alternating warm and cold advections of air masses and changing wind directions, which drive coastal ocean and estuarine waters to oscillate quasi-periodically. Quantifying water transport under these meteorological oscillations between inland waterways and the coastal ocean helps interpret land-ocean interactions and the impact of migrating weather systems. The challenge lies in the difficulty of obtaining continuous, long-term direct measurements of transport due to logistical constraints. Here, we apply a method to determine the meteorological tide-induced volume transport of water through an intensive survey, correlating transport values measured by a boat-mounted ADCP with vertically averaged velocities from a bottom-mounted ADCP, which recorded a much longer time series. The correlation is then used to compute transport over the period of the bottom-mounted ADCP deployment. Observations were conducted at Belle Pass, Port Fourchon. The transport data revealed the impact of weather systems, including four cold fronts. A model of volume transport, accounting for rotary cold front wind variations, was applied, where both along-channel and along-coastline wind components contribute to the remote wind effect, leading to a more complex response to passing weather systems. The local wind effect is much smaller than the remote wind effect, and transport is primarily controlled by water level fluctuations resulting from open boundary input. Finally, the channel orientation relative to the coastline is found to be critical in determining both the magnitude and phase of the transport.

## 1. Introduction

Water as a carrier of dissolved and suspended materials provides the momentum for the global land-ocean exchange. The transport of water and waterborne materials such as suspended sediment, carbon, nutrients, and pollutants is determined by the hydrodynamics in the rivers, estuaries, and ocean (Lin et al., 2022; Yang et al., 2023; Baumas and Bizic, 2024). To estimate the transport of water and waterborne materials it requires the quantification of water flow velocity and the concentration of materials of interest (Defendi et al., 2010), among other parameters, depending on the specific substance of interest.

The transport of mass ($T_m$) per unit time of a given material is an integration of flux ($\mathbf{F}$) across an area (such as a vertical cross section in the mouth of an estuary or a tidal channel) through which the transport process occurs, in which the flux is determined by the products of velocity and concentration ($c$) of the substance:

$$\mathbf{F} = c\mathbf{v} \quad (1)$$

$$T_m = \int_A c\mathbf{v} \cdot d\mathbf{A} \quad (2)$$

where $A$ is the area of the cross section and $d\mathbf{A}$ is the area element vector perpendicular to the cross section. Here we assume that $c$ uses a unit of mass per unit volume and equation (2) gives a mass transport per unit time (e.g., kg/s). If $c$ uses a unit of ppt (parts per thousand), to get the total mass transport in unit time, $c$ in (2) must be replaced by $c\rho$ to compute the mass transport in unit time in which $\rho$ is density. For simplicity, without losing its essence, here we only discuss volume transport (or discharge) of water in unit time ($T_V$):

$$T_V = \int_A \mathbf{v} \cdot d\mathbf{A} \quad (3)$$

a commonly used quantity in studies of water circulation (e.g., Talley et al., 2011; Guo et al., 2022; Shin et al., 2022).

For most applications, the present-day measurements of water flow velocity are best done by acoustic Doppler current profilers (ADCP), using technology developed since the 1980's (Teledyne, 2011) and continuously perfected till today. To measure the transport of water across a given section in the river, tidal channel, estuary, or a bay, one can use an ADCP mounted on a moving platform (such as a boat) across the section (e.g. Wang et al., 2003). Most ADCPs come with a software package that can output transport directly. Alternatively, especially for wider channels, a harmonic analysis of velocity field can be done first (Candela et al., 1992; Münchow, 2000). The time series of transport can then be reconstructed (Wang et al., 2003; Codiga and Aurin, 2007). Here we only limit the discussion to narrow channels, and we use the transport reported from the instrument. This method assumes that during the time the moving platform makes a complete transect the velocity field does not change significantly. This is not a problem if the cross section is "narrow", or if the time it takes for one transect is much smaller than the timescale of the velocity variation (such as a tidal cycle). To resolve the transport change over time, the measurements must be done with continuous occupation of the transect, which is costly and difficult to continue for long-term observations. In addition, when a severe weather system approaches a sampling site, it is unlikely a safe operation of a continuous vessel-based survey can be implemented.

An alternative approach is to use a bottom mounted ADCP, or an array of ADCPs. The advantage of bottom mounted ADCPs is that they can survive most weather events, except under an extreme situation such as a direct hit of a major tropical cyclone. Since the velocity field is rarely uniform across a section and even several bottom mounted ADCPs might be too sparse to reliably reconstruct the cross-sectional transport using data from bottom mounted ADCP alone.

A third method is to simultaneously use a vessel-based ADCP transecting a cross section over a period during which velocity field has a significant temporal variation and a bottom-mounted ADCP deployed at a point along the transect for a much longer time. A statistical relationship can be examined between the two datasets during the vessel-based measurements and determine if the flow velocity from the bottom mounted ADCP is correlated with the total transport (Li et al., 2018; Li and Boswell, 2022). If this is the case, the velocity data from the bottom mounted ADCP can be used to rescale to a transport time series.

In this work, we apply this method to a tidal channel - Belle Pass at Port Fourchon, Louisiana, U.S.A. The next section describes the study site and method, followed by Section 3 on the application and results, Section 4 with discussion on the impact of weather systems, and Section 5 for the summary.

## 2. Study Site and Method

### 2.1. Study site

The study site (Fig. 1) is located on the southern Louisiana coast at Belle Pass (south of Port Fourchon), which is the last section of Bayou Lafourche toward the ocean (Fig. 1a). It is a slightly curved channel of ~ 250 m wide. This channel meanders northward toward inland for tens of kilometers along the wetland in the region, with some bifurcations along the main channel. At about 33 km along the main channel, there is the Leno Theriot Lock at ~29°20’34”N, 90°14’46” W. This lock appears to be mostly open but does present a restriction. At the ocean end near its mouth, there are a couple of 0.7 – 0.8 km long rock jetties. The study site is about 2 km from the coastal ocean. This area is dominated by diurnal micro-tides (Kantha, 2005) with a maximum tidal range of ~ 0.6 m. As a result, the weather influence on hydrodynamics and water transport is relatively more important compared to the tidal effect. Since the adjacent estuaries and bayous are crucial to fish nursery, there has been an interest in wind induced water transport and its impact to larvae transport (e.g. Norcross and Shaw, 1984; Kupchik, 2014).

Port Fourchon is a major port of the northern Gulf of Mexico with many industries, providing a land base to support the offshore oil and gas industry. It serves more than 90% of the deepwater oil production in the Gulf of Mexico. It is estimated that there are more than 500 oil and gas production units in the ocean within 40 miles of Port Fourchon. Because of its strategic location for the ocean navigation, the importance to the energy industries and local economy, and its low elevation in a wetland area that is affected by abnormal relative sea-level rise and repeated impact of winter storms (Zhao et al., 2022), the hydrodynamics and water transport through Port Fourchon is of great interest but there has been little hydrodynamic and hydrographic data from the region.

### 2.2. Method

The objective of the study is to use a combination of observations from a short-term vessel-based survey and a relatively long-term bottom deployment to determine the water transport time series induced by meteorological forcing – the meteorological tide (Murty, 1984). Here the term “meteorological tide” is defined as the water level variation and related water flow velocity variation caused by weather systems moving over the region under the large-scale atmospheric circulation. Meteorological tide has its weather origin (Murty, 1984; Li et al., 2020) and thus is very different from its astronomical counterpart (the astronomical tide).

The method to determine the water volume transport time series under the influence of meteorological tide involves two steps. The first step is the use of a vessel to conduct repeated measurements of water transport across the channel. At the same time, a bottom mounted current meter is used to continuously record the flow velocity time series. This allows the possibility of establishing a correlation between the velocity from the bottom deployed current meter and the vessel measured total transport. The rationale is that when the total transport is large, the velocity from a point measurement should also be relatively large and vice versa. The second step is, once a correlation is established through a regression analysis, the velocity time series from the bottom mounted current meter recorded for a much longer time, during which no cross-channel direct measurement of total water transport from a boat is available, can be used to compute the volume transport for the rest of the time series. This method has been used in, e.g., a tidal river (Hu et al., 2023), a tidal channel (Li et al., 2018; Weeks et al., 2018), and an arctic lagoon (Li and Boswell, 2022) and is proven to be useful. The reconstructed water transport can then be used to study the impact of weather systems on the transport by analyzing the variability in response to the passing weather systems.

## 3. Observations

### 3.1. Data from Bottom Mounted Instruments

An ADCP was deployed at the bottom of Belle Pass to record current velocity and hydrographic data for a month and half. A 1200 kHz RDI ADCP was deployed roughly 4 m below the surface in Belle Pass at (29° 5'53.19"N, 90°13'18.18"W), started sampling at about 1630 UTC, March 30, 2010. The vertical bin size was set to be 0.5 m sampling at 50 pings per ensemble with 15-minute ensemble average intervals. This continued until 1930 UTC, May 11, 2010, before the vessel based ADCP survey when the ADCP was retrieved, and data downloaded; and the ADCP was immediately redeployed. The redeployment had a different ensemble average interval of 2 minutes, with 0.2 m vertical bins and 50 pings per ensemble. The higher sampling rate was set to examine correlation of velocity with the transport measurements from the vessel based ADCP. The second deployment with higher sampling rate and higher vertical resolution lasted until the end of the vessel based ADCP survey at around 2312 UTC May 13, 2010. This makes the length of deployment to be about 44 days.

### 3.2. Vessel based Continuous ADCP Measurements

An 8.5-m long twin-engine catamaran was used to carry a 1200 KHz ADCP along a rectangular box in the channel of Belle Pass repeatedly (Fig. 1c). The ADCP measured the vertical profiles of the three-dimensional velocity components (u, v, w) at ~0.52 Hz (~1.9 seconds) for 25 hours. The survey was conducted between 1422 UTC, May 12 and 1522 UTC, May 13, 2010. The rectangle is about 220 m wide and 500 m long. The maximum water depth in the center of the channel was about 9 m while the depth on the banks was about 1-2 m. The ADCP was installed on a stainless-steel pole mounted on the port side of the boat at about 0.4 m below the surface. The vessel speed was maintained at about 4-5 knots (2-2.5 m/s). Over the 25-hour period, the cross channel transects were repeated more than 100 times.

### 3.3. Weather Data

Meteorology data (surface wind speed, wind direction, barometric pressure, air temperature, dew point temperature, and relative humidity) sampled at 20-minute intervals was obtained from an ASOS station at Fourchon (NOAA station ID GAO, station name Galliano, located at 29.44482°N, 90.26112°W, Fig. 1b). Weather maps from the National Weather Prediction Center of NOAA were also used in the analysis. These weather maps were obtained by ground stations nationwide at 3 hourly intervals. Particularly, we used the weather maps of four cold front passages of the region around April 8, April 25, May 3, and May 8 (Table 1), respectively, during the deployment for the analysis of the influence of meteorological tides. Additional meteorological and oceanographic time series data was obtained from an offshore station managed by the WAVCIS lab, Louisiana State University – the CSI 6 station (the ST 52B oil platform) at (28° 52.077' N, 90°29.466'W), about 35 km southwest of the study site on the Louisiana continental shelf at a water depth of about 20 m. Data from CSI 6 include air temperature, water temperature, barometric pressure, and wind speed and direction.

# 4. Results

## 4.1. Velocity and Depth Variations

The temporal variation of the velocity field was recorded along the transect over the 25-hour period. As an example, the time series of surface velocity components at the center (29.0982° N, -90.222° W) of the northern transect, within a circle of 10 m in diameter, is shown in Fig. 2a. The variation is mainly tidal: by using a diurnal (with a period of $P$=24 hour) and a semi-diurnal (with $P$=12 hour) tidal constituents and a mean, a harmonic analysis yields a fit to the measurements with an $R^2$ value of 0.80 for the east velocity component and 0.89 for the north velocity component (figure omitted). Since the main channel has an angle of ~ 64° from the latitude line, a rotation of the coordinate system with that angle is done to convert the wind velocity vector to the along and across channel velocity components. Figure 2a shows the time series of rotated velocity components. A harmonic analysis fit gives an $R^2$ value of 0.92 for the along channel velocity. Similarly, the water depth variation exhibits tidal variations - Fig. 2b shows the depth variation at the center of the transect. Apparently, tidal signal is mainly diurnal, with a tidal range of about 0.5 m and an $R^2$ value of 0.92.

## 4.2. Regression between Velocity and Transport

In addition to the velocity field, among the recorded variables from the instrument of the vessel based ADCP is the volume transport of the water relative to the bank (edge of water). The time series shows tidal variations which are correlated with the depth averaged along channel velocity measured from the bottom deployed ADCP. Figure 3a shows the comparison between the two. For convenience of visualization, we multiplied the depth-averaged along channel velocity ($u$) from the bottom deployed ADCP by 2115. A linear regression shows that the transport is $\alpha = 2114.8$ times of the mean along channel velocity $u$. The $R^2$ value of the regression is 0.97 with a $p$ value of 0 (Fig. 3b). This result verifies the implied null hypothesis that the total cross sectional volume transport of water is linearly correlated with the depth-averaged along channel velocity. If we apply this factor to the entire time series of the observed velocity from the bottom deployed ADCP, a time series of transport is obtained (Fig. 4a).

The temporal variations include tidal and subtidal signals. The subtidal time series is obtained by an application of a low pass filter using the Fourier filter (O'Haver, 2023; Li, 2023) with a cutoff frequency of 0.6 cycle per day. The weather induced variations are best visualized by the low pass filtered data (red line in Fig. 4a). The relatively large variations are associated with the change in weather conditions. For example, when there is a cold front passage, there is a relatively large variation in the subtidal transport (Fig. 4a). Typically, prior to the passage of a cold front, wind is from the southern quadrants, and the flow is from the ocean to the channel, while after the frontal passage, the flow reverses its direction to drain the channel or bay (Li et al., 2020). These are shown in Figure 4 around the times of the frontal passages. After the passage of a cold front, a rebound of the water level, velocity, and transport may follow, as shown in Fig. 4a. The rebound was immediate for the first cold front around April 8 although those for the second and third cold fronts appear to be less abrupt. This application has provided a time series of volume transport for the entire time of the bottom deployed measurements and provided quantification of the response to the passing weather systems.

The wavelet transform (Fig. 4b) shows that during the first and second cold fronts, there was an increase in high-frequency energy above the diurnal tidal oscillations, extending beyond the random fluctuations. This influence was short-lived, as indicated by the narrow yellow zones in time (approximately within a day). In contrast, the influence of the cold front in the low-frequency bands was present across all cases, with a broader temporal impact spanning both the pre- and post-front periods. However, the correlation with the timing of the cold front passages was less pronounced.

## 5. Discussion

The time series of transport allows us to study the weather system induced hydrodynamic response that determines the transport variations, thereby providing an approach to study the impact of weather and climate. In the following, we further evaluate the role of weather by discussing the meteorological data and associated transport variability.

### 5.1. Wind Variations

Wind direction during the bottom ADCP deployment between March and May varied significantly. Overall, the wind direction was dominated by southeasterly (Fig. 5a). During a cold front event, the wind reverses its direction when the front passes. In general, the pre-frontal wind is a warm advection in the lower atmosphere with relatively wet and warm maritime tropical (mT) air mass (Hsu, 1988) from the subtropical ocean (Fig. 6). Wind is from the southern quadrants before the passage of a cold front. After the passage of the front, wind is mostly from the northern quadrants with cold advection when a relatively cold and dry air mass moves from the north into the region (Roberts et al., 1989) with an increasing barometric pressure (Fig. 6). The details of temporal variation of wind can vary appreciably because cold fronts can have several factors varying independently with several degrees of freedom so that the temporal variation in wind can be convoluted. These factors include the orientation of the cold front, the shape and track of the cold front, the track and evolution of the associated extratropical cyclone (which can be hundreds or even thousands of kilometers away in the north), the moving speed of the air masses, and the intensities of the low- and high- pressure systems associated with the front.

In this study, the four cold fronts had different temporal variations in wind speed and direction as demonstrated by the wind rose plots. Figures 5a, 5b, and 5c are wind rose plots for the first three cold fronts, respectively. The last cold front occurred around May 8 was rather weak. Each wind rose plot includes two days of wind data, from one day before to one day after the passage of each cold front. Prior to the first cold front, wind was southeasterly, followed by northwesterly before changing to roughly northerly. This results in a wind rose plot with multi-modal form (Fig. 5b). Figures 6a and 6b are weather maps before and after the time the cold front passed the study site, respectively, showing the change in wind direction, in addition to the warm and cold advections as shown by the change in surface temperature (upper left number in red) and dew point temperature (lower left number in green) by the station model data. The wind rose of the second cold front showed a dominant wind from the southwest. One day before the arrival of the cold front at the study site, wind started from southeasterly, then southerly, and southwesterly before switching to roughly westerly (Fig. 5c). The third cold front had a predominantly southeasterly because of a strong prefrontal southeasterly wind although the post frontal wind also had northwesterly, northerly, and northeasterly components, they were much weaker in magnitude (Fig. 5d).

Wind velocity time series (Fig. 7a) from the ASOS station (GAO) indicates that during the first cold front, wind velocity components are the weakest, compared to the second and third cold fronts. The southerly wind reaches its maximum just shortly before the arrival of the cold front, which corresponds to the inward transport of water (Fig. 4a). After the passage of a cold front, the reversed wind direction corresponds to the resultant outward transport (Fig. 4a). The wind data from CSI 6 station and that of the ASOS data from GAO station are quite consistent with the former having a slightly greater magnitude (figure omitted).

It should be noted that cold fronts in May can be much weaker than those in late fall, winter, and early spring. A weak cold front is sometimes shown by a much weaker northerly wind after the passage of the front. This is because the Arctic air mass has weaker momentum when the air temperature gradient is reduced as the summer gets closer. As a result, the central northern Gulf of Mexico during this period often experiences fewer cold frontal passages and the dominant wind is southeasterly. When there is a cold front passage, although the southeasterly pre-frontal wind can be strong (like in this case for the 3rd front on 3 May), the post frontal wind can be very weak or even non-existent as the high-pressure system behind the fronts is weaker.

In our case, as seen in Figure 7, the wind after the 3$^{rd}$ frontal passage was indeed very weak (circled by blue dashed oval on Fig. 7a), while the pre-frontal southeasterly wind was particularly (dashed ovals in Fig. 7a).

The weather map from NOAA for 1800 UTC on 3 May 2010 (not included in the figures) shows that behand the cold front, there was a weak and small-scale high-pressure system centered at the border of northeast Louisiana and western Mississippi. Because of the proximity of the center of the high-pressure, the study site is more like in the warm advection region, rather than cold advection region (unlike most of the winter cold front conditions with the high-pressure system further to the northwest). This makes the wind behind the cold front not having typical northerly winds. This is corroborated by Fig. 7a. For more typical situations, previous studies in this region are referred to Roberts, et al. (1989), Walker and Hammack (2000), and Li et al. (2019).

### 5.2. The warm and cold advections

The air temperature and dew point temperature time series (Fig. 7b) from the ASOS weather station GAO also show typical response during cold front passages. The prefrontal warm advection of the air mass resulted in an increase of the air temperature. Before the passage of the cold front, both the air temperature and dew point temperature reached their maximum values as the mT air mass reached its maximum strength which is sometimes associated with precipitation. At the passage of the cold front, the barometric pressure reached its minimum (Fig. 7c). The offshore data from CSI 6 station allowed the comparison between the water and air temperature changes during cold front events (Fig. 7d). While the signals of warm and cold advections of the air affected the air temperature over the shelf water much the same way to air over land, the water temperature change was less obvious except during the first cold front. The water temperature had a steadier increase during the time as a seasonal variation. The correlation coefficient between water temperature and air temperature is 0.89. The increasing temperature trend is aligned with the seasonal change (getting into the summer) so that both water and air temperatures were increasing during the study period. The cold front caused some perturbations and brief drops in temperature, but the signal was rather weak and buried in the noise, particularly compared to the seasonal change.

### 5.3. Comparison between Remote and Local Wind Effects

The response of volume transport to temporally varying wind involves both local and remote effects. Garvine (1985) provided a study using an analytic model in which the flow field in a channel is solved for a single frequency component and the wind is assumed rectilinearly varying with time at the given frequency of a weather event. The remote wind effect is assumed to be that of water level variation at the open boundary and purely caused by the Ekman transport along the coastline. In Feng and Li (2010), the wind vector was modified to allow a rotary form of variation to simulate the cold front wind. In Garvine (1985), the orientation of the channel can be slanted relative to the coastline, while in Feng and Li (2010), the channel is perpendicular to the coastline. In the case of Belle Pass, both the channel and coastline are tilted at an acute angle from the east and some modifications can generalize the model to include such a geometric setup. In addition to allowing the channel and coastline to be slanted, we use transport instead of velocity in the momentum and continuity equations. In addition, in view of the recent work of Huang et al. (2024) that the water level at the mouth of the bay is not only affected by the Ekman effect but also by the onshore wind component, we include an additional contribution to the remote wind effect. Thus, the remote wind effect is a combination of influence from both the alongshore and onshore wind components.

We use the depth-averaged shallow water equations with the following variables and parameters: time ($t$), wind stress component ($\tau_{ax}$) along the channel (in the $x$ direction), water elevation ($\zeta$), depth-averaged flow velocity along the channel ($u$), mean water depth ($h$), water density ($\rho$), width ($D$), and linear frictional coefficient ($\beta$),

$$\frac{\partial u}{\partial t} = -g\frac{\partial \zeta}{\partial x} + \frac{\tau_{ax}}{\rho h} - \frac{\beta u}{h} \quad (4)$$

$$\frac{\partial \zeta}{\partial t} + \frac{\partial hu}{\partial x} = 0 \quad (5)$$

where $\beta$ is derived from a Fourier series expansion of the quadratic bottom friction (Parker, 1984):

$$\beta = \frac{8C_d U_0}{3\pi} \quad (6)$$

in which $C_d$ and $U_0$ are the bottom drag coefficient and estimated magnitude of the along channel velocity, respectively. Since the volume transport of water ($T$) is

$$T = Dhu \quad (7)$$

The equations for transport $T$ can be derived from (4) and (5):

$$\frac{\partial T}{\partial t} = -gDh\frac{\partial \zeta}{\partial x} + \frac{D\tau_{ax}}{\rho} - \frac{\beta T}{h} \quad (8)$$

$$D\frac{\partial\zeta}{\partial t}+\frac{\partial T}{\partial x}=0 \tag{9}$$

As in Garvine (1985), we look for a periodic solution of the idealized model under a periodic wind stress (simulating the cold front wind with clockwise rotation). In such a case, the along channel (Fig. 8) wind stress component will be sinusoidal and can be expressed as

$$\tau_{ax}=Re\{\tau_0 e^{i\omega t}\} \tag{10}$$

Here $\tau_0$ is the amplitude of the wind stress, and $\omega$ the angular speed of an oscillation due to the meteorological tide. The symbol $Re$ is an operator taking the real part of the complex number within the braces. The solution for transport and water elevation can be expressed as

$$T=Re\{T_0(x)e^{i\omega t}\} \tag{11}$$

$$\zeta=Re\{A(x)e^{i\omega t}\} \tag{12}$$

where $T_0$ and $A$ are the amplitudes of the water volume transport and water elevation, respectively. Substituting (10)-(12) into (8) and (9) yields

$$\left(i\omega+\frac{\beta}{h}\right)T_0=-gDh\frac{dA}{dx}+\frac{D\tau_0}{\rho} \tag{13}$$

$$i\omega DA+\frac{dT_0}{dx}=0 \tag{14}$$

The boundary conditions are

$$T_0|_{x=L}=0 \tag{15}$$

$$A|_{x=0}=A_0 \tag{16}$$

The solution can be found as follows

$$\zeta=Re\left\{\left(\frac{\tau_0}{\alpha\rho gh}\frac{\sin(\alpha x)}{\cos(\alpha L)}+A_0\frac{\cos[\alpha(x-L)]}{\cos(\alpha L)}\right)e^{i\omega t}\right\} \tag{17}$$

$$T=Re\left\{\left(\frac{\gamma}{\alpha^2}\left[1-\frac{\cos(\alpha x)}{\cos(\alpha L)}\right]+A_0\frac{i\omega D}{\alpha}\frac{\sin[\alpha(L-x)]}{\cos(\alpha L)}\right)e^{i\omega t}\right\} \tag{18}$$

where

$$\alpha^2=\frac{\omega^2}{gh}\left(1-i\frac{\beta}{\omega h}\right) \tag{19}$$

$$\gamma=-i\frac{D\tau_0\omega}{\rho gh} \tag{20}$$

It is verified that the equations (17) and (18) satisfy the equations (8) and (9) and the boundary conditions. The local wind effect is represented by the terms with $\gamma$ or $\tau_0$, while the remote wind effect is shown by terms involving $A_0$.

In this solution, the local wind effect ($T_L$) contributing to the transport is:

$$T_L = Re\left\{-i\frac{D\tau_0}{\rho\omega\left(1-i\frac{\beta}{\omega h}\right)}\left[1-\frac{\cos(\alpha x)}{\cos(\alpha L)}\right]e^{i\omega t}\right\} \quad (21)$$

whereas the remote wind effect ($T_R$) contributing to the transport is

$$T_R = Re\left\{A_0\frac{iD\sqrt{gh}}{\sqrt{1-i\frac{\beta}{\omega h}}}\frac{\sin[\alpha(L-x)]}{\cos(\alpha L)}e^{i\omega t}\right\} \quad (22)$$

In Garvine (1985), the remote wind effect is expressed as

$$A_0 = aE \quad (23)$$

in which $E$ is the Ekman transport $E = \tau/f$, $f$ is the Coriolis parameter, $\tau$ is the wind stress component along the coastline, and $a$ is a remote wind coefficient. In Feng and Li (2010), the remote wind effect is given a 90° phase difference with the local wind stress component, simulating the rotary cold front wind. With the additional contribution to the remote wind effect from onshore wind component to the water level at the mouth, equation (23) can be modified to be

$$A_0 = a\tau_{al0} + b\tau_{an0} \quad (24)$$

where $\tau_{al0}$ and $\tau_{an0}$ are the complex amplitude of wind stress in the direction of the coastline $l$ (positive in the downcoast direction or Kelvin wave direction, Fig. 8) and in the direction $n$ perpendicular to coastline (positive onshore), respectively. The parameters $a$ and $b$ are the coefficients similar to the remote wind coefficient in Garvine (1985). Given that the downcoast direction of the coastline has an angle of $\varphi$ (Fig. 8) with the latitude line, while the channel has an angle of $\theta$ with the latitude line, and assume that the wind stress vector is rotary and clockwise rotating with an oscillation frequency of $\omega$, as assumed in Garvine (1985) and Feng and Li (2010), then its east ($\tau_{aE}$) and north ($\tau_{aN}$) components will be:

$$\tau_{aE} = \tau_a\cos(\omega t) \quad (25)$$

$$\tau_{aN} = -\tau_a\sin(\omega t) \quad (26)$$

where $\tau_a$ is the magnitude of wind stress. We have the following expression of the wind stress components in the directions of *n* and *l*, respectively:

$$\tau_{an} = -\tau_a\sin(\omega t + \varphi) \quad (27)$$

$$\tau_{al} = -\tau_a\cos(\omega t + \varphi) \quad (28)$$

We can prove the following relations (Fig. 8):

$$\tau_0 = \tau_a e^{i\theta};\ \tau_{an0} = i\tau_a e^{i\varphi};\ \ \tau_{al0} = -\tau_a e^{i\varphi} \quad (29)$$

In Garvine (1985), the remote wind effect is introduced by the open boundary elevation related to the Ekman flux by $\zeta(0) = \alpha\,\tau/f$, in which $\alpha$, $\tau$, and $f$ are a constant coefficient, wind stress along the coastline, and Coriolis parameter, respectively. The value for $\alpha$ used in Garvine (1985) is $5\times10^{-3}$ $cm^3\cdot dyn^{-1}\cdot s^{-1}$ or $5\times10^{-4}$ $m^2\cdot s\cdot kg^{-1}$, which is equivalent to $a = \alpha/f = 50/7.18\sim6.96$ $m^2\cdot s^2\cdot kg^{-1}$. In comparison, in Feng and Li (2010), the Coriolis parameter is $f$=7.18 × $10^{-5}$ $s^{-1}$ and the value for $\alpha$ is $8 \times 10^{-5}$ $m^2 \cdot s \cdot kg^{-1}$, which is equivalent to a value for $a = \alpha/f = 8/7.18 \sim 1.1$ $m^2 \cdot s^2 \cdot kg^{-1}$, in (24). To estimate the value for $b$ in equation (24), we estimate the water setup across a shelf of 20 km width with an averaged depth of 15 m, the following relationship should hold:

$$\Delta\zeta \sim \frac{\tau_a \Delta x}{\rho g h} \sim 0.13\tau_a \qquad (30)$$

So, an estimated value for $b$ is on the order of 0.13 ($m^2 \cdot s^2 \cdot kg^{-1}$). In our computation, the length of the channel is defined to be $L$=100 km. This value is crude for our system and somehow subjective: although the channel can go up to 100 km, there is a manmade constriction at about 35 km upstream of the study site in the Bayou Lafourche. It is however still connected to upstream. The channel width is $D$ = 250 m, average water depth $h$ = 6 m, and the transect of ADCP measurements is at $x = 2.5$ km from the mouth. We further define the cold front period $P = 50$ hours (~ 2 dt in Table 1 for the second cold front or for the average condition), air density $\rho_a = 1.29$ (kg/m$^3$), water density $\rho = 1025$ (kg/m$^3$), air drag coefficient $C_{da} = 1.22 \times 10^{-3}$, bottom drag coefficient $C_{db} = 2.5 \times 10^{-3}$. With these parameters, the solution is computed with four different scenarios (Table 2):

(1) We use the parameters consistent with those of Feng and Li (2010), e.g., $a$=1.1 $m^2 \cdot s^2 \cdot kg^{-1}$, but assumed a channel orientation angle $\theta$ = 64.4° (the estimated value for Balle Pass) and a coastline angle of $\varphi$=33.5° (Fig. 9a).

(2) We use a larger value for $a$ (5 $m^2 \cdot s^2 \cdot kg^{-1}$), $\theta$ = 64.4°, and $\varphi$=33.5° (Fig. 9b).

(3) Same as Figure 9a but the channel is oriented north-south, or $\theta$ = 90°, and the coastline is oriented west-east, or the coastline angle $\varphi$=0° (Fig. 9c).

(4) Same as Figure 9b but the channel is oriented north-south, or $\theta$ = 90°, and the coastline is oriented west-east, or the coastline angle $\varphi$=0° (Fig. 9d).

In all these scenarios (Figure 9), the results for the time series of one cycle of transport variations for remote and local wind contributions and the total transport are shown. Obviously, for the first and third scenarios, both local and remote wind effects are underestimated. The local effect is determined (not adjustable) while the remote wind effect is determined through the parameters $\alpha$ (Garvine, 1985; Feng and Li, 2010) or $a$ and $b$ (this paper). Figure 9b is a computation of the solution with an increased $a$ value of 5 $m^2 \cdot s^2 \cdot kg^{-1}$, which resulted in a total transport magnitude comparable to the observations of ~ 600 m$^3$/s (Fig. 4 low pass filtered – red curve). This value for the remote wind effect $a = 5$ $m^2 \cdot s^2 \cdot kg^{-1}$ is ~28% smaller than that of Garvine (1985) ($a = 6.96$ $m^2 \cdot s^2 \cdot kg^{-1}$). Indeed, the remote wind effect contributes much more than the local wind effect.

These results also show that the orientation of both the channel and the coastline plays a crucial role in determining the magnitude and phase of the water volume transport induced by the meteorological tide (comparing Fig. 9b and 9d).

## 6. Summary

By combining vessel-based observations with bottom-mounted instrument measurements, we obtained a strong linear correlation ($R^2$ = 0.97) between the total volume transport of water from the vessel-based ADCP data over a 25-hour period and the depth-averaged along-channel velocity measured from a bottom mounted ADCP. This correlation allows us to compute a time series of volume transport over the entire duration of the longer-term bottom based observations of velocity profiles. The variability in volume transport during this period can then be examined after low pass filtering to remove tidal oscillations, for the impact of meteorological tide, specifically cold fronts, including both prefrontal and post frontal periods. Over the 44-day deployment, four cold fronts influenced the study site. This method enables the quantification of transport influenced by these weather systems without requiring direct measurement of the total transport from the vessel for the entire 44 days.

By modification to the models of Garvine (1985) and Feng and Li (2010), the volume transport of water is solved with forcing from a clockwise rotating wind stress, simulating the wind variation of a cold front passage. The model considers the effect of a slanted channel with an arbitrary angle with the coastline. The contributions to the transport from remote and local wind effects are separated. In the remote wind effect, a contribution from onshore wind is added, which however proves to be only a minor modification, compared to the Ekman flux effect. The transport in the Belle Pass is largely dominated by the water level setup and set down due to the Ekman transport through the connectivity with the coastal ocean at the open boundary of the tidal channel (Garvine, 1985). Whereas the contribution to transport from the local wind effect by the wind stress acting on the water inside the Belle Pass is much smaller (only ~ 10% of that from the remote wind effect). For a channel such as Belle Pass, the meteorological tide induced water transport is mainly determined by the coastal ocean response, and not by the wind stress acting on the interior water body of the tidal channel. This study also shows that the orientation of the channel relative to the coastline has a critical influence on both the magnitude and phase of the transport induced by meteorological tide. Although the model results are sensitive to the length of the channel and the remote wind effect coefficients ($a$ and $b$), the conclusion that remote wind effect is more important in a system like Belle Pass, shallow and narrow, is quite robust.

## Acknowledgements

This study is partially supported by the National Science Foundation (Award Number 1736713) and NOAA (NA21NOS0120092) through GCOOS. The vessel-based survey and the bottom-mounted ADCP deployment were conducted with the assistance of participants, including captains, technicians, and students from the Coastal Studies Institute at Louisiana State University. Helpful review comments from anonymous reviewers were also appreciated.

## Data Availability Statement

Data for this work is available at https://repository.lsu.edu/oceanography_coastal_wavcis/8/.

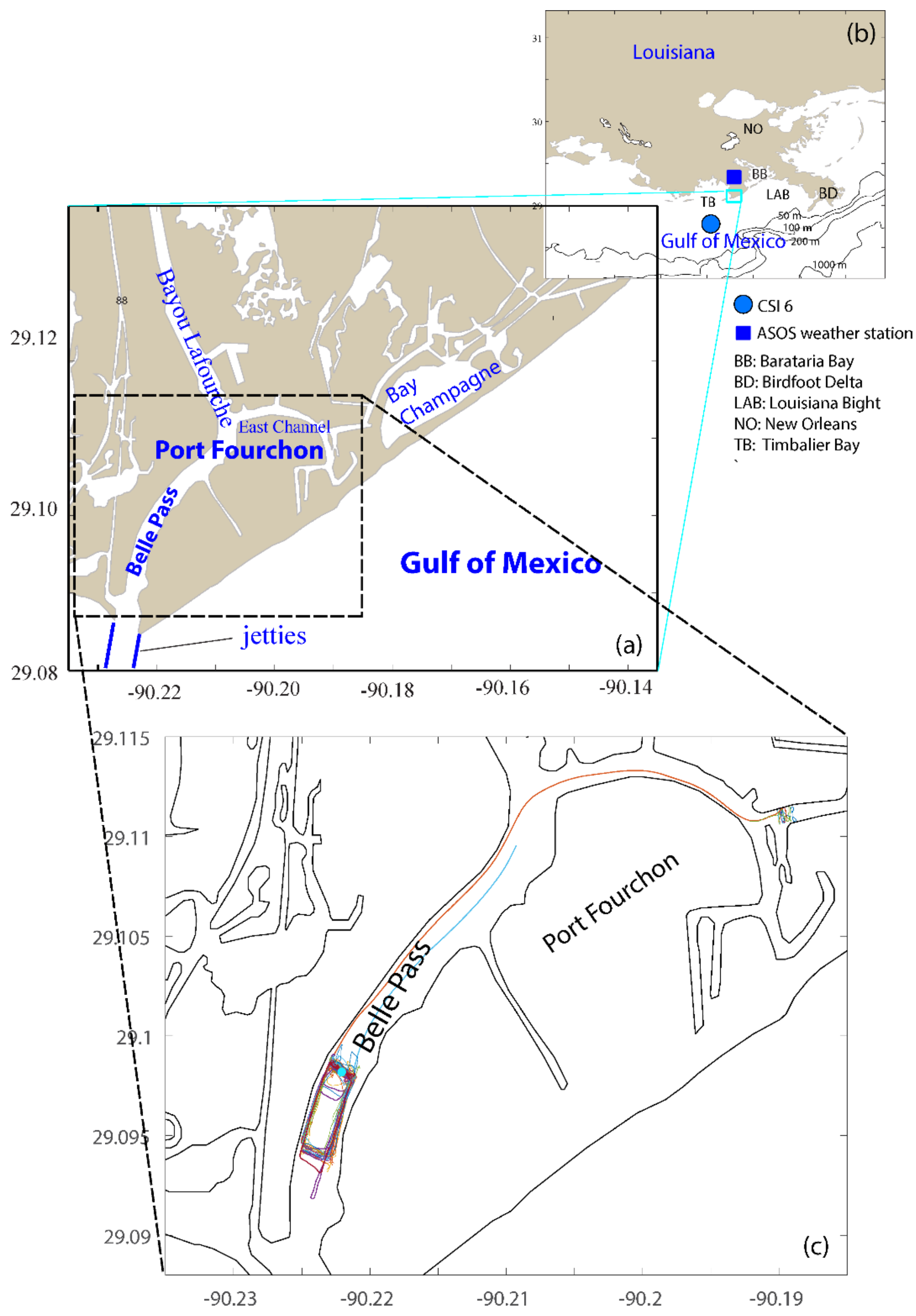

**Figure 1.** Study site. (a) Port Fourchon region with the tidal channel Belle Pass, in which the vessel based tidal cycle survey and longer-term deployment were made. (b) A larger area around the study site with the ASOS weather station shown by the blue rectangle. (c) Ship-track from the continuous tidal cycle survey using a 26-ft catamaran at Belle Pass, Port Fourchon, Louisiana on May 12-13, 2010. Location of bottom mounted ADCP was shown by the cyan dot.

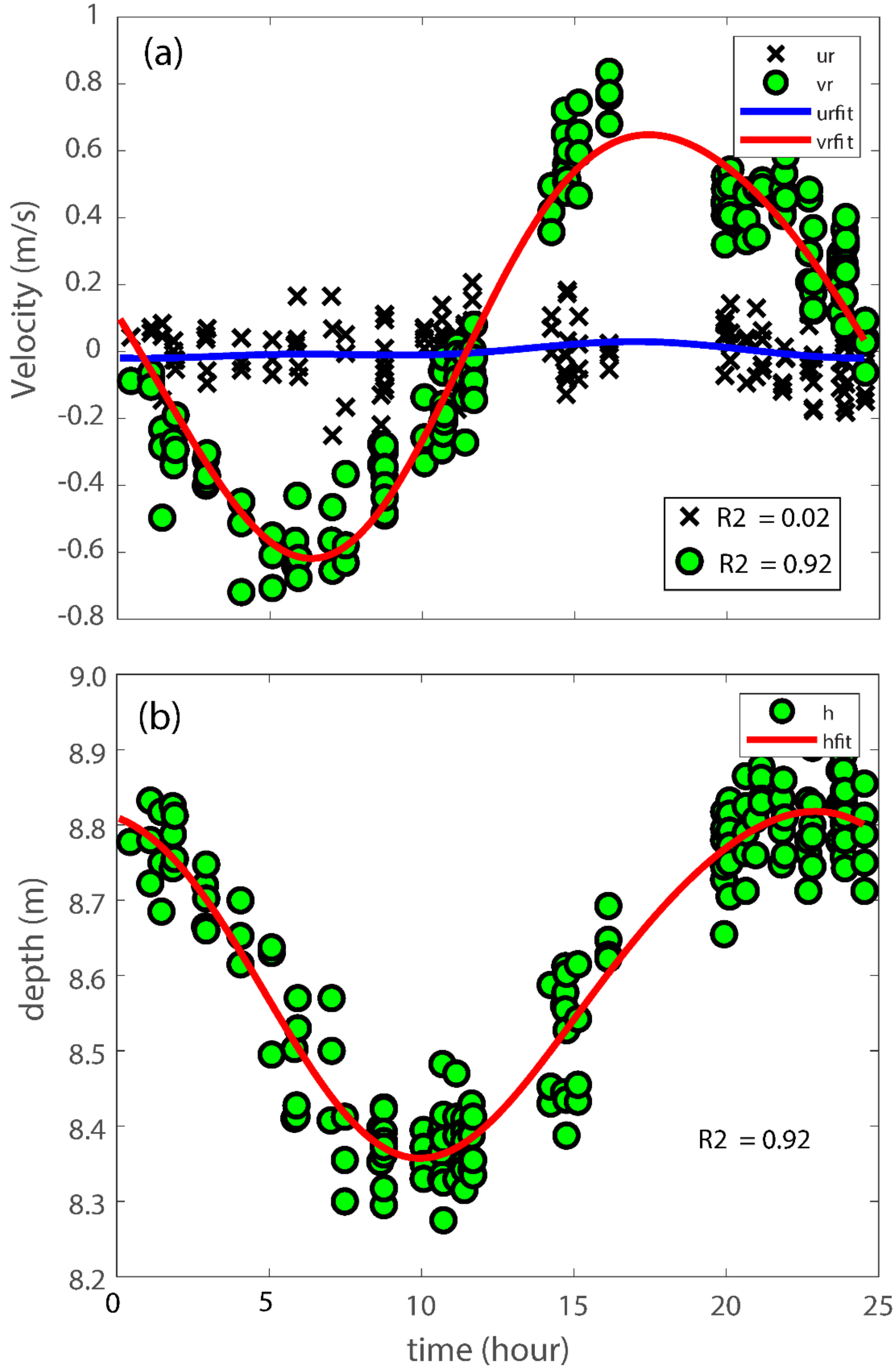

**Figure 2**. Data from the vessel based ADCP. (a) Surface along-channel (green circles) and cross-channel (crosses) current velocity components at the center of the channel and fitted time series including semidiurnal and diurnal constituents. (b) Water depth variation at the center of the channel and fitted time series including semidiurnal and diurnal constituents.

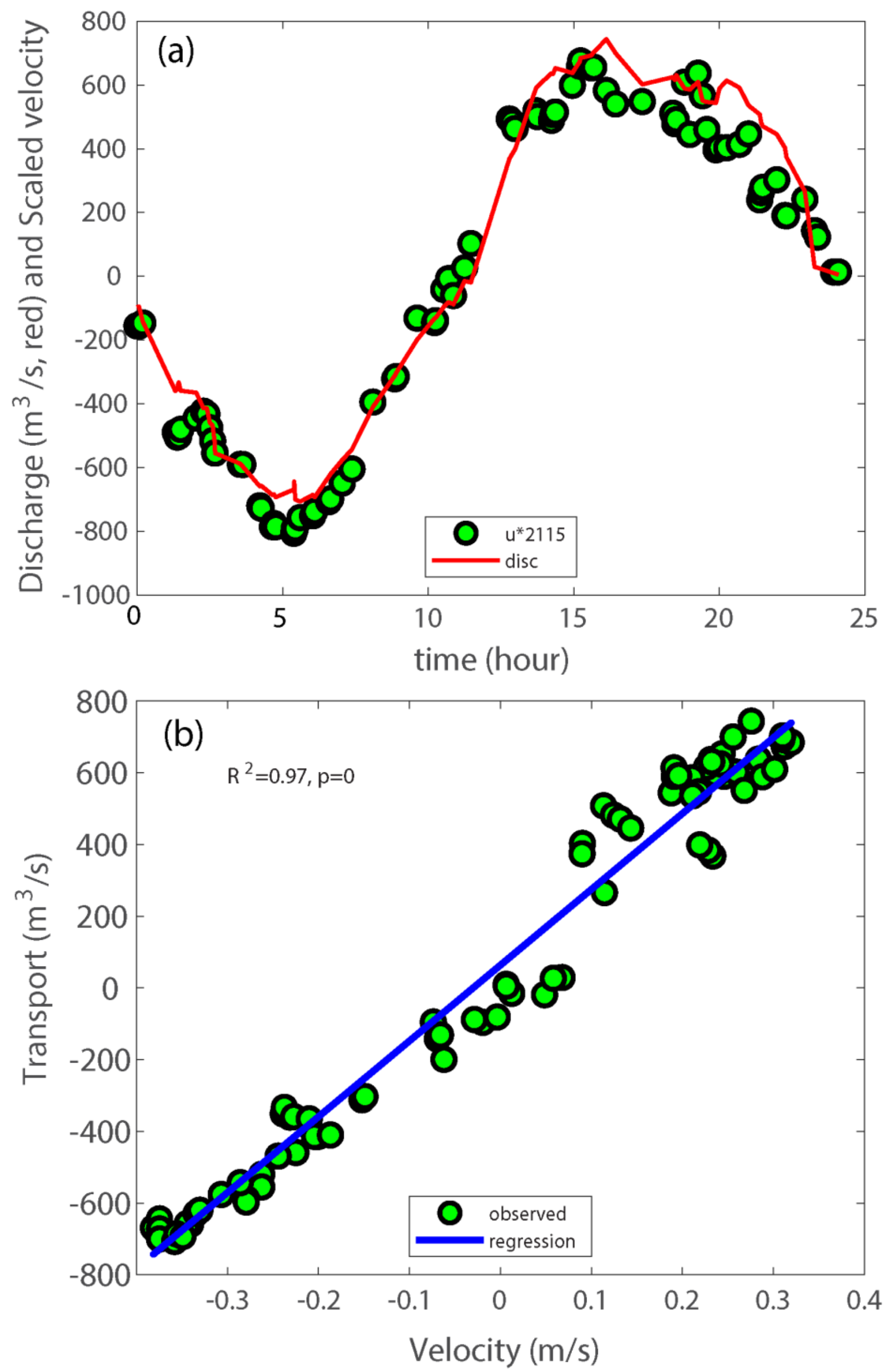

**Figure 3.** Correlation between depth averaged velocity from the bottom ADCP and the integrated transport from the vessel based ADCP. (a) Time series comparison between the depth averaged velocity from the bottom ADCP multiplied by 2115 and the integrated transport from the vessel based ADCP. (b) Regression between the depth averaged velocity from the bottom ADCP and the integrated transport from the vessel based ADCP.

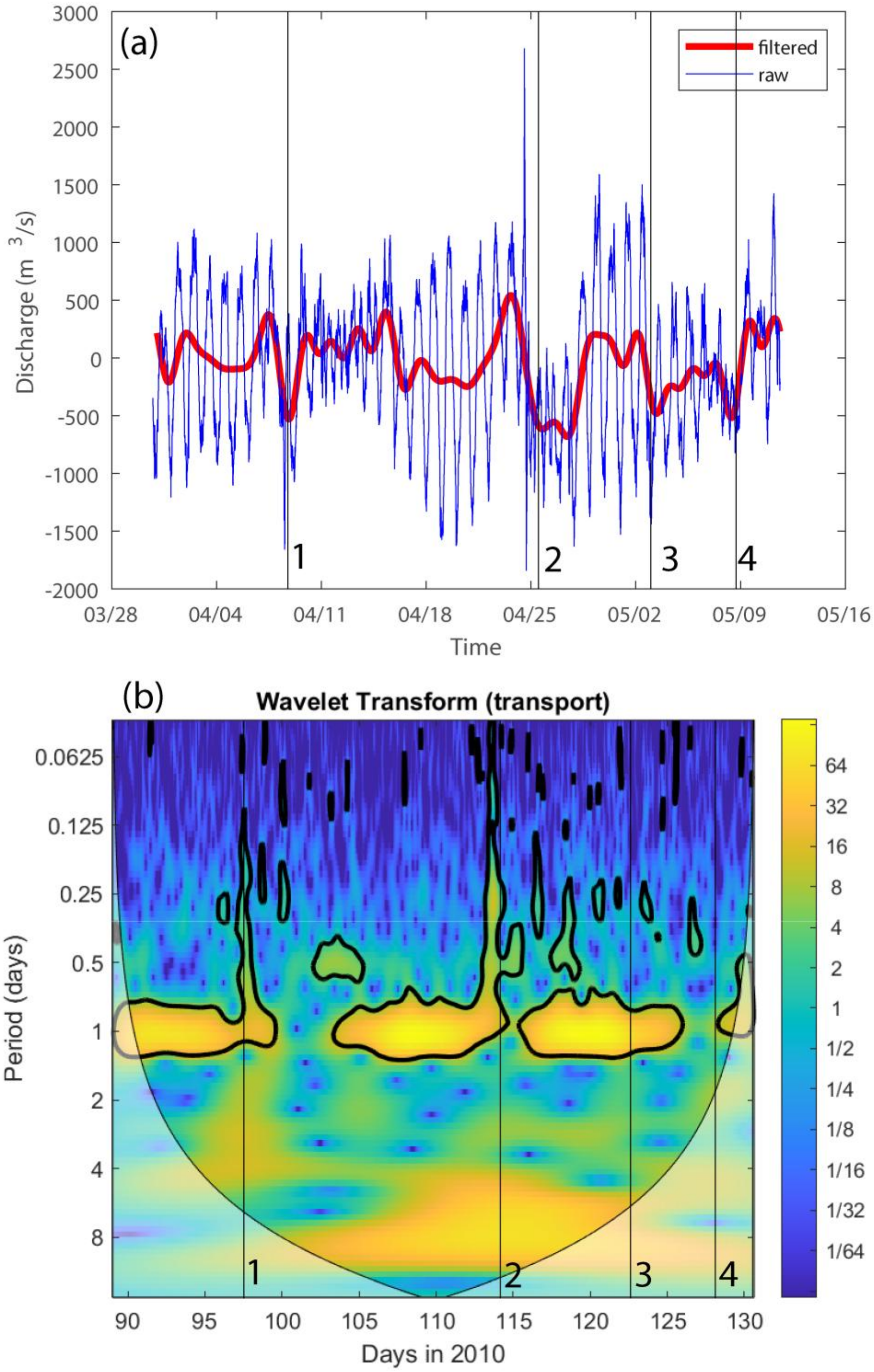

**Figure 4**. Transport time series and its wavelet transform. (a) Regression derived total discharge (volume transport) at Belle Pass (blue line) and its low pass filtered time series (red line). (b) The wavelet transform of the transport. The vertical bars indicate the time of rough estimate of cold front passage of the study site.

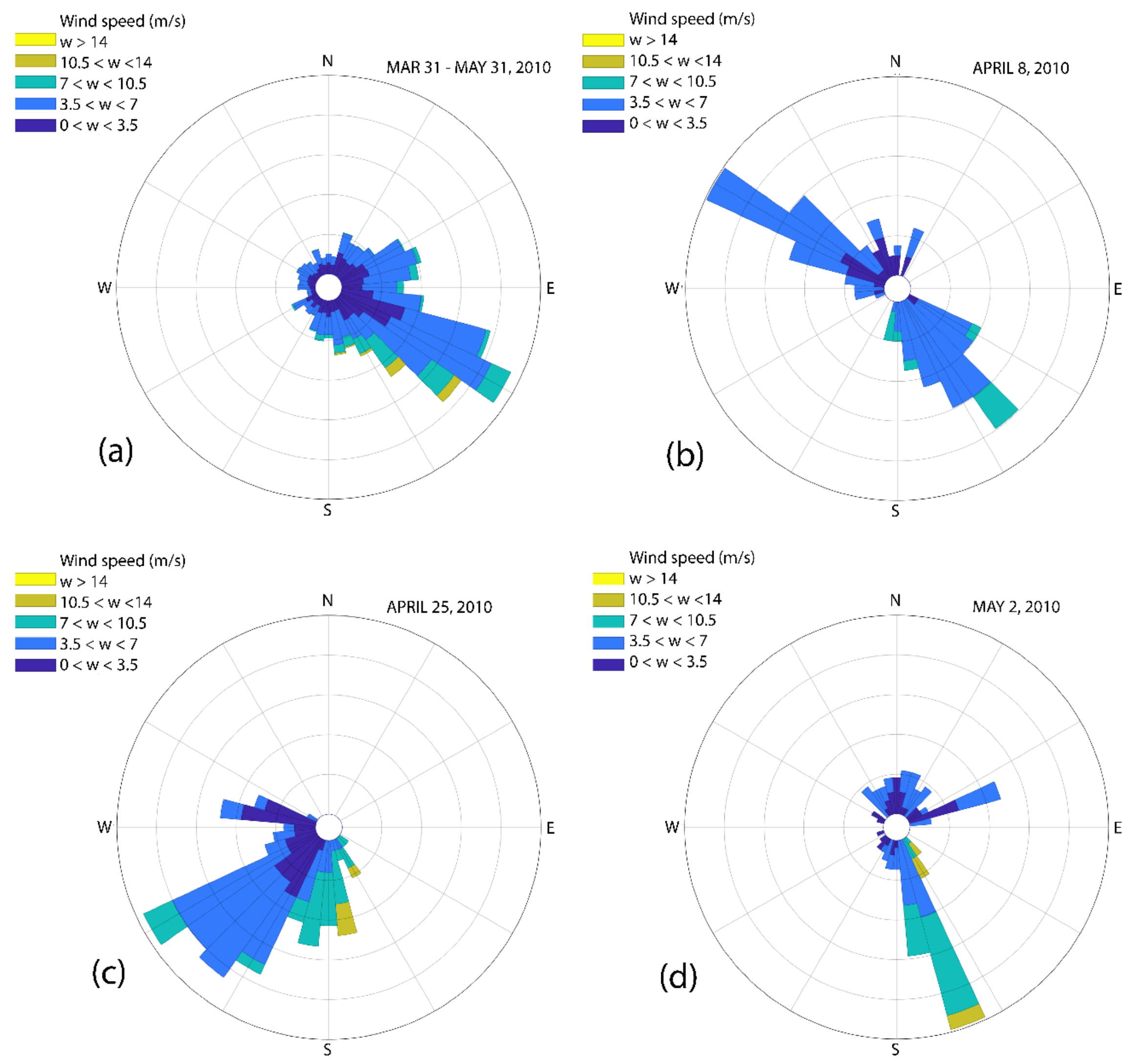

**Figure 5.** Wind rose plots. (a) The overall wind rose for the entire observational period. (b) Wind rose plot for the two days centered at the time of passage of the first cold front around April 8. (c) Wind rose plot for the two days centered at the time of passage of the second cold front around April 25. (d) Wind rose plot for the two days centered at the time of passage of the third cold front around May 2.

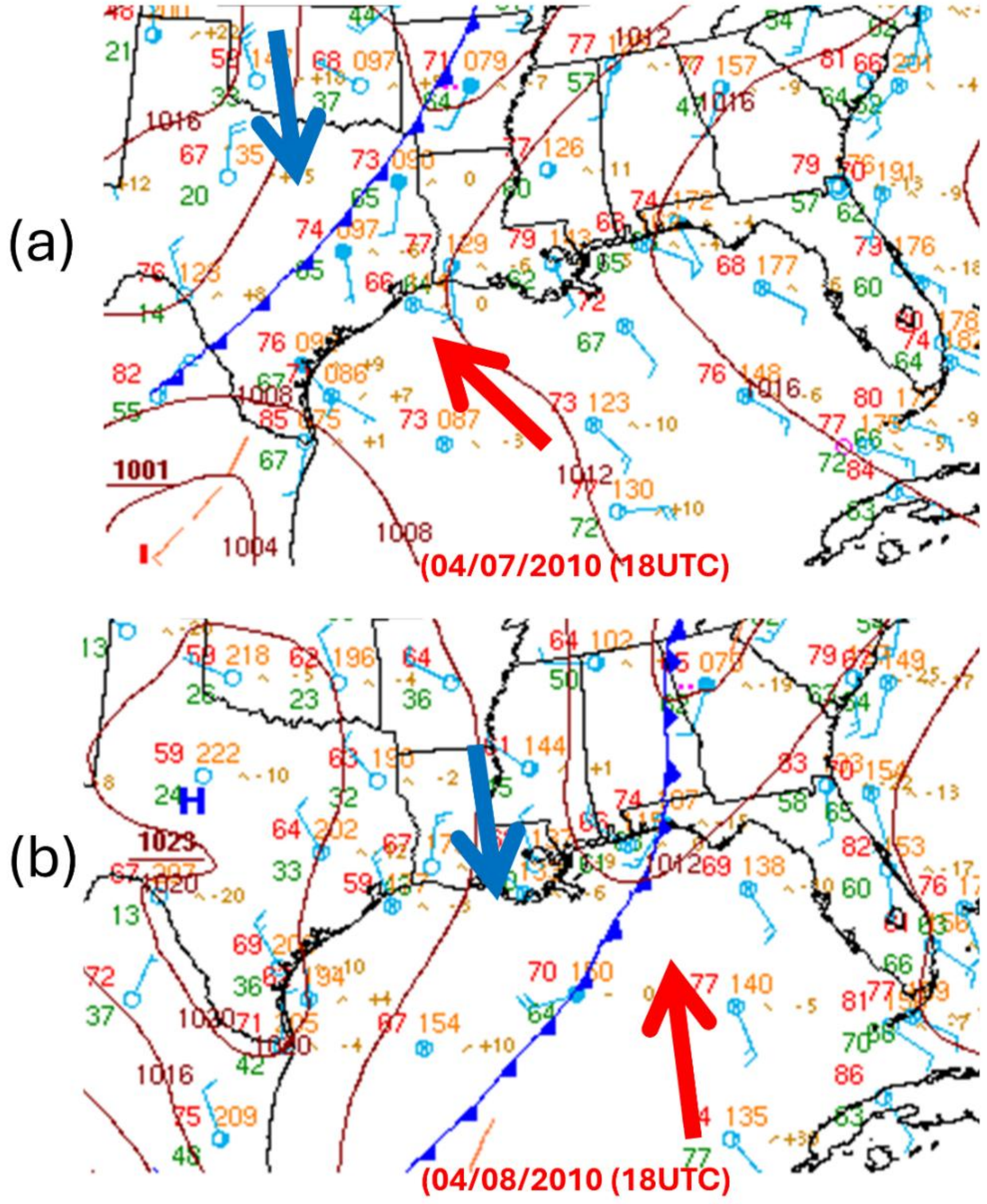

**Figure 6.** Weather maps before (a) and after (b) the passage of the first cold front during the deployment. The red arrows indicate regions of warm advection, and the blue arrows indicate cold advection.

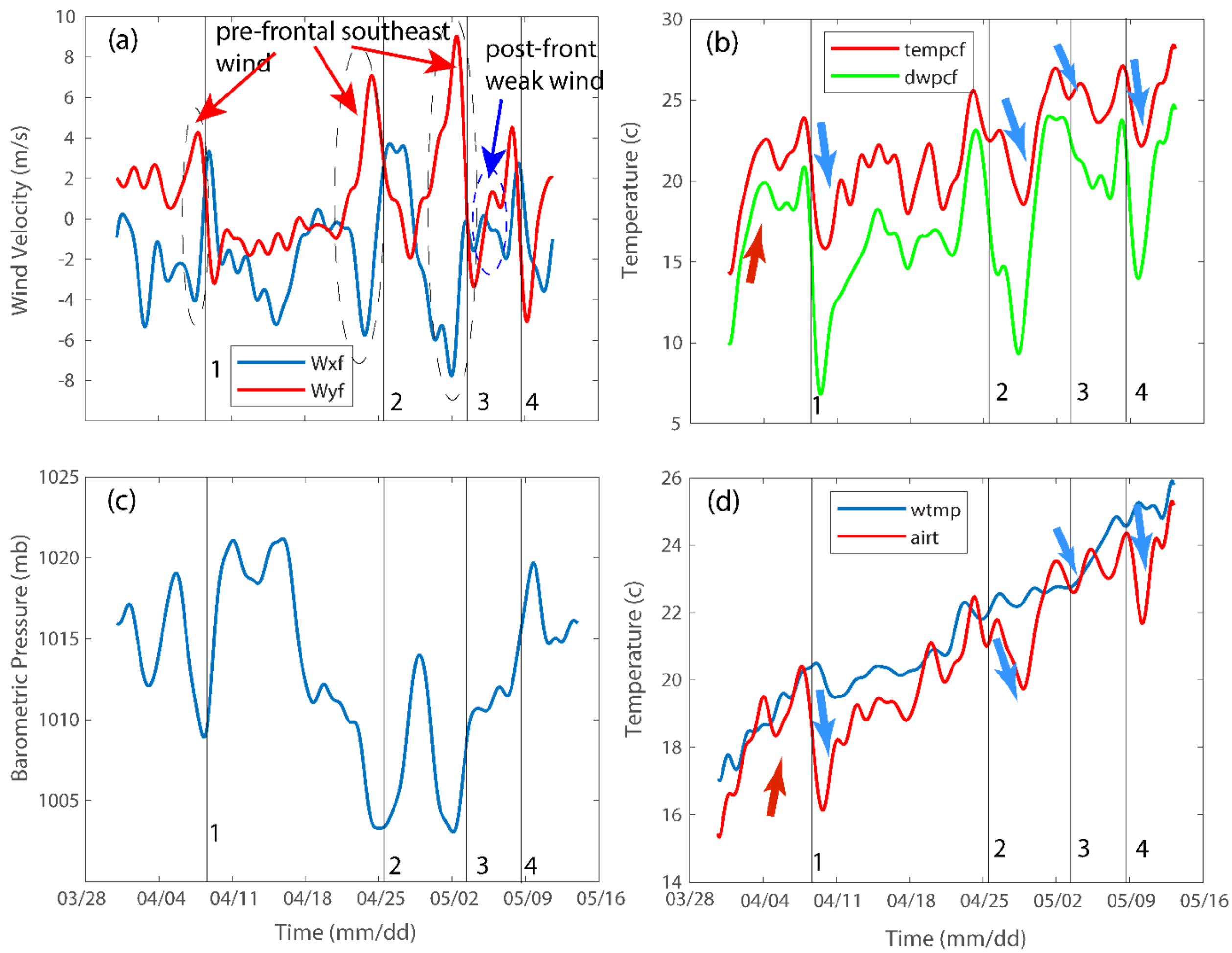

**Figure 7**. Low pass filtered observations (time series) from the ASOS station GAO and CSI 6 (Fig. 1b). (a) Wind velocity components (east component – blue and north component – red) from GAO. (b) Air temperature and dew point temperature from GAO. (c) Barometric pressure from GAO. (d) Water temperature (blue) and air temperature (red) from CSI 6. The vertical bars indicate the time of rough estimate of cold front passage of the study site. Red and blue arrows indicate examples of the effect of warm and cold advections, respectively.

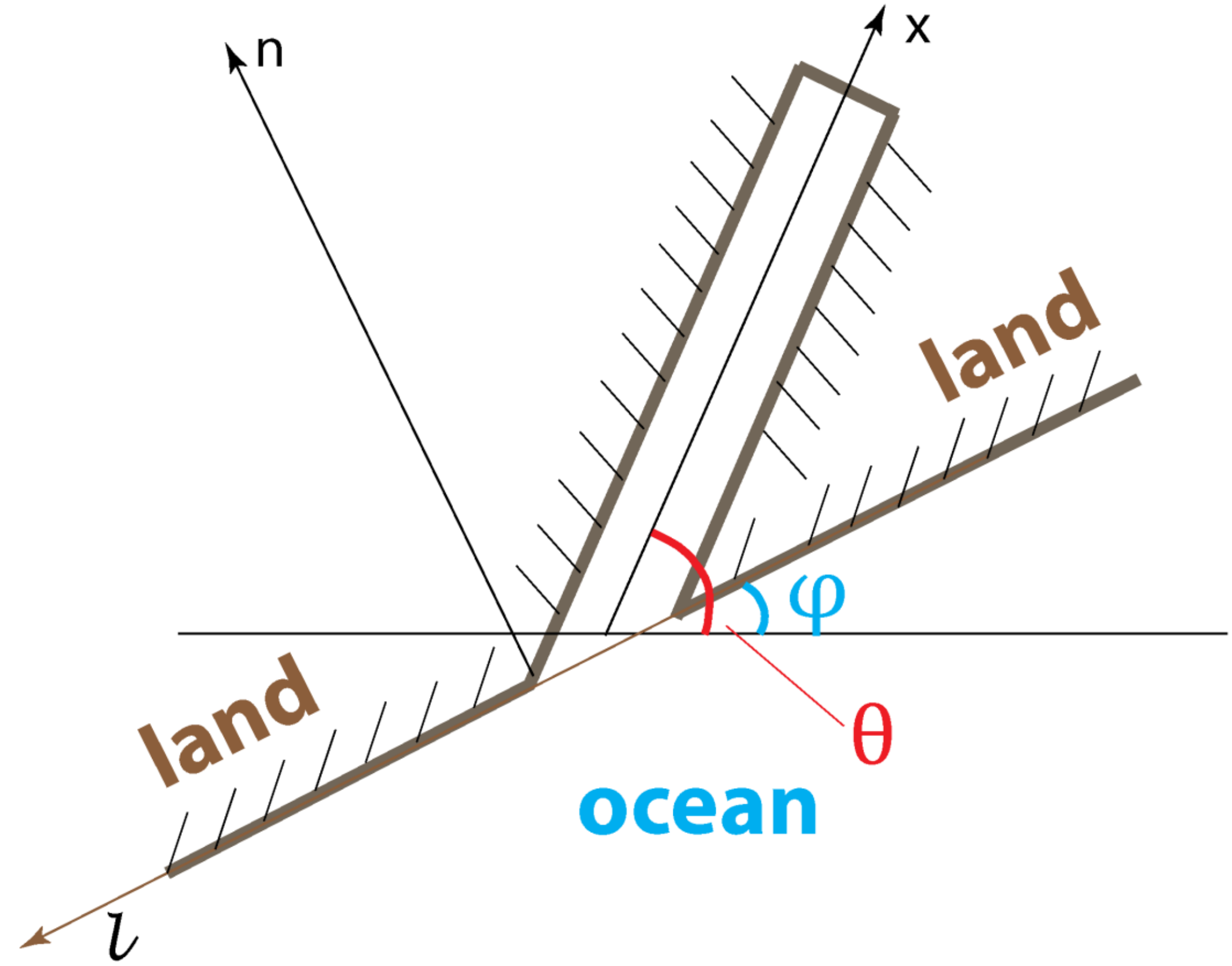

**Figure 8**. Model estuary. The coastline has an angle of $\alpha$ with the latitude line. The tidal channel has an angle of $\beta$ with the latitude line. The channel is along x, the downcoast direction is along $\boldsymbol{l}$, while the normal direction of the coastline is along n.

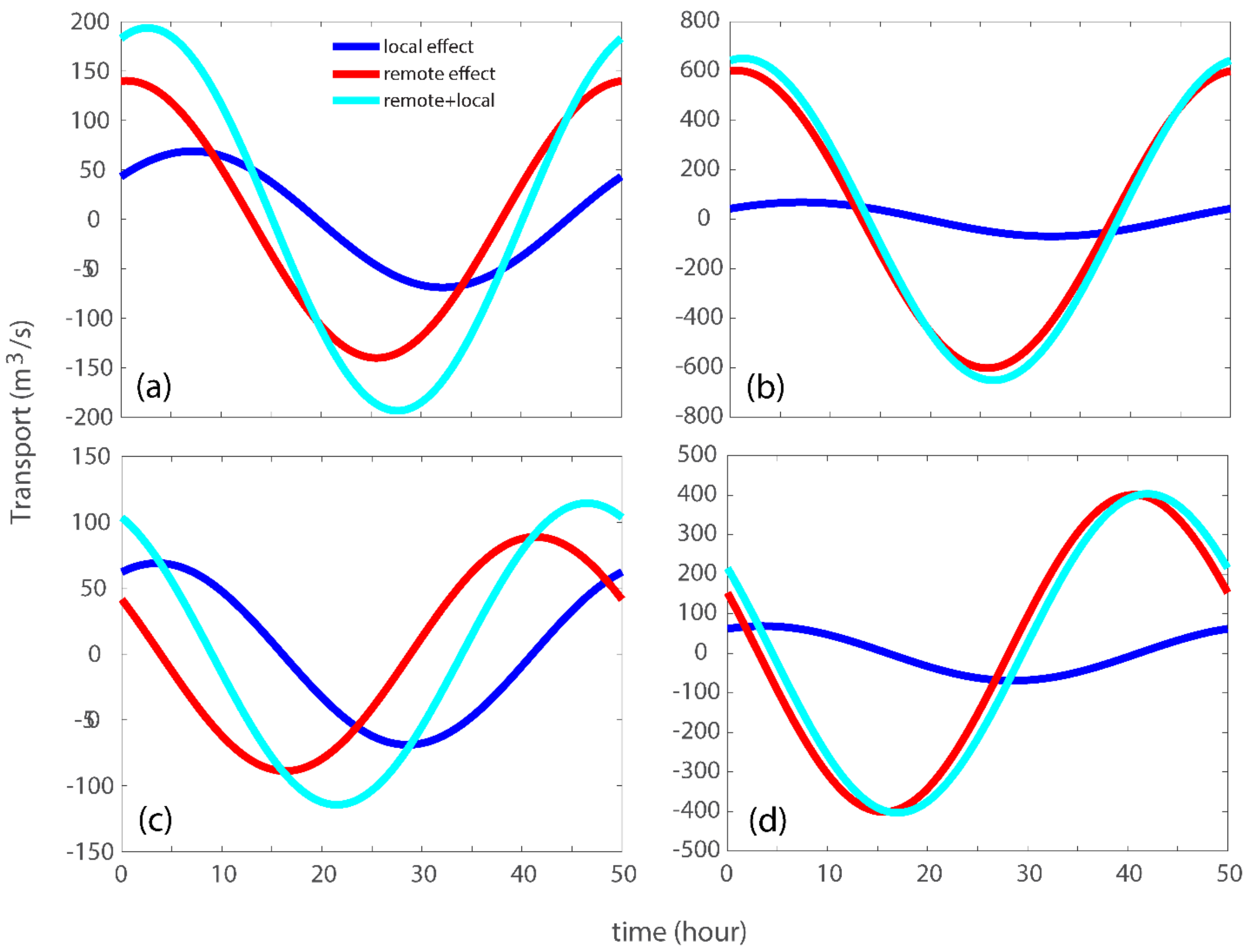

**Figure 9.** Model output for the remote and local wind effect induced transport and total transport (Table 2). (a) Using the same $\alpha$ as in Feng and Li (2010) but for a channel with an orientation angle of 64 degree and a coastline orientation of 33 degrees (Fig. 8). (b) Same as (a) but using 4.55 times of the $\alpha$ as in Feng and Li (2010). (c) Same as in (a) but with a channel perpendicular to the coastline. (d) Same as (b) but with a channel perpendicular to the coastline (Table 2).

**Table 1.** Cold fronts during the study period

| No. | Time of cold front entering LA | Time of cold front passing study site | Time of cold front leaving LA | dt |
|---|---|---|---|---|
| 1 | 0 UTC April 8 | 12 UTC April 8 | 15 UTC April 8 | 15 hours |
| 2 | 12 UTC April 24 | 12 UTC April 25 | 15 UTC April 25 | 27 hours |
| 3 | 6 UTC May 2 | 12 UTC May 3 | 18 UTC May 3 | 36 hours |
| 4 | 22 UTC May 7 | 18 UTC May 8 | 19 UTC May 8 | 21 hours |
| | | | Average dt (approximately) | 25 hours |

**Table 2.** Parameters used in the model with cold front rotary wind

| Parameter name | Case 1 (Fig. 9a) | Case 2 (Fig. 9b) | Case 3 (Fig. 9c) | Case 4 (Fig. 9d) |
|---|---|---|---|---|
| Period (hour) | 50 | 50 | 50 | 50 |
| Width D (m) | 250 | 250 | 250 | 250 |
| Length (km) | 100 | 100 | 100 | 100 |
| Depth (m) | 6 | 6 | 6 | 6 |
| Wind speed amplitude (m/s) | 8 | 8 | 8 | 8 |
| Air density $\rho_a$ (kg/m$^{-1}$) | 1.23 | 1.23 | 1.23 | 1.23 |
| Water density $\rho$ (kg/m$^{-1}$) | 1025 | 1025 | 1025 | 1025 |
| Surface drag coefficient $C_{da}$ | 1.22×10$^{-3}$ | 1.22×10$^{-3}$ | 1.22×10$^{-3}$ | 1.22×10$^{-3}$ |
| Bottom drag coefficient $C_{db}$ | 2.5×10$^{-3}$ | 2.5×10$^{-3}$ | 2.5×10$^{-3}$ | 2.5×10$^{-3}$ |
| Location of output $x$ (distance into the channel, km) | 2.5 | 2.5 | 2.5 | 2.5 |
| Remote wind effect $a$ | 1.1 | 5 | 1.1 | 5 |
| Remote wind effect $b$ | 0.13 | 0.13 | 0.13 | 0.13 |
| Velocity magnitude $U_0$ (m/s) | 0.5 | 0.5 | 0.5 | 0.5 |
| Channel orientation $\theta$ (°) | 64 | 64 | 90 | 90 |
| Coastline angle $\varphi$ (°) | 33 | 33 | 0 | 0 |